\documentstyle[12pt]{article}

\newcommand{\nl}{\nonumber \\}

\def\beq{\begin{equation}}
\def\eeq{\end{equation}}
\def\beqa{\begin{eqnarray}}
\def\eeqa{\end{eqnarray}}

\begin{document}


\vskip .3cm

\centerline{\Large \bf Duality in supersymmetric gauge theories} 

\vskip 1cm

\centerline{\bf Paolo Di Vecchia, 
                   \footnote{e-mail: DIVECCHIA@nbivms.nbi.dk} }
\centerline{\sl NORDITA}
\centerline{\sl Blegdamsvej 17, DK-2100 Copenhagen \O, Denmark}

\vspace{1cm}
\centerline{ {\it Lectures given at the 24 ITEP Winter School (February 1996)}}
\vspace{1cm}







\begin{abstract}
In these lectures we discuss various aspects of gauge theories with extended
$N=2$ and $N=4$ supersymmetry that are at the basis of recently found exact 
results. These results include the exact calculation of the
low energy effective action for the light degrees of freedom in the $N=2$
super Yang-Mills theory and the conjecture, supported by some checks, that the
$N=4$ super Yang-Mills theory is dual in the sense of Montonen-Olive. 
\end{abstract}

\vspace{1cm}

\section{Introduction}
\label{intro}

In the last few years a number of exact results have been
obtained both in four-dimensional supersymmetric gauge theories and in
supersymmetric string theories in various dimensions. They are all based on 
one hand on supersymmetry and on the other hand on the exploitation of a 
duality  symmetry that is the generalization of the duality present in free 
electromagnetism and that allows one to relate theories, which at the
first sight look very different, in much the same way  as in two dimensions
Sine-Gordon is related to the Thirring model.

In these lectures we will not discuss string theories, but we will limit 
ourselves
to the study of supersymmetric Yang-Mills theories with extended supersymmetry
$N=2$ and $N=4$.  In particular we will discuss in some detail the background
that is needed for obtaining the exact results found by Seiberg and 
Witten~\cite{SEIWI}
for $N=2$ supersymmetric Yang-Mills theory, that will be also shortly reviewed,
and for formulating the Montonen-Olive~\cite{MONTOLI} duality conjectured for 
the $N=4$ super Yang-Mills theory.  

Several recent reviews~\cite{OLIVE,HARVEY,BILAL} have appeared in the last 
year. In particular the beautiful review by David Olive~\cite{OLIVE} 
has strongly inspired these lectures. 

The plan of the lectures is as follows. 

In section (\ref{dua}) we will 
discuss  duality in electromagnetism and the Dirac quantization condition.
In section (\ref{mono}) the 't Hooft-Polyakov magnetic monopole and 
the Julia-Zee dyon solutions in the Georgi-Glashow model are discussed in some 
detail. These classical solutions  correspond to particles in the quantum 
theory. This implies that the particle spectrum does not only contain the 
particles 
corresponding to the fields present in the classical Lagrangian and having
a constant or small mass in the weak coupling limit, but also contains 
other particles corresponding to the soliton solutions and having a big mass
in the weak coupling limit. Those additional particles are also required by
certain duality properties that extend the duality of  electromagnetism 
discussed in section (\ref{dua}). The discussion of the soliton solutions in 
sect. (\ref{mono}) brings us in section
(\ref{MOdua}) to the Montonen-Olive duality conjecture. After its formulation
it became
very soon clear that this duality property cannot be satisfied in the
Georgi-Glashow model where the quantum corrections invalidate the conclusions
based on  semiclassical considerations. The theories that have a chance to
realize it were those in which the semiclassical properties are not
destroyed by quantum corrections and those are the supersymmetric gauge 
theories. That is why in sect. (\ref{repre}), as an introduction to them, we 
discuss the representations
of supersymmetry algebra with and without central charges and in sect. 
(\ref{supe}) we construct the supersymmetric Yang-Mills actions from
dimensional reduction from $D=10$. In sect. (\ref{semi}) we present the 
semiclassical analysis of the $N=2$ super Yang-Mills theory, that can also
be easily extended to the $N=4$ case, showing that 
this theory has magnetic monopoles and dyon solutions as the Georgi-Glashow
model discussed in sect. (\ref{mono}) and presents all the main features
that brought to the Montonen-Olive duality conjecture for the Georgi-Glashow
model. In particular in the cases with extended supersymmetry the mass formula
for the BPS states is not just a property of the classical theory, but is a 
consequence of the supersymmetry algebra and as long
as supersymmetry is not broken, is  an exact quantum formula. In section 
(\ref{centra}) we write the supersymmetry algebra for the case $N=2$ and
we derive from it an expression for the mass of the BPS states.
From the previous considerations it appears that both the $N=2$ and $N=4$
super Yang-Mills theories are perfectly good candidates for the realization
of the Montonen-Olive duality conjecture. Actually it turns out that the $N=2$
theory cannot verify it. We discuss this issue in sect. (\ref{Neq24}) where
we bring various arguments that select the $N=4$ theory as the best
candidate for the Montonen-Olive duality conjecture. In sect. (\ref{SZ})
we discuss the Schwinger-Zwanziger quantization condition and from it
and other very general assumptions we show that the electric and magnetic 
charges of the physical states must lie on a two-dimensional lattice. 
Then, using the mass formula for the BPS-states, we give a characterization
of the single particle stable states. In sect. (\ref{RIMO}), following very
closely Ref.~\cite{OLIVE}, we riformulate the Montonen-Olive duality conjecture
adapted to the $N=4$ super Yang-Mills theory and we show that the various
formulations are related to each other by the action of the modular group 
$SL(2,Z)$. The last three sections provide a short summary of the beautiful
paper by Seiberg and Witten~\cite{SEIWI} where the low energy effective 
action for the
light degrees of freedom is constructed. In particular in sect. (\ref{global})
we describe the global parametrization of the moduli space, in sect 
(\ref{singu})
we discuss its singularities and finally in sect. (\ref{exsolu}) we give the
explicit solution found in Ref.~\cite{SEIWI}.

\section{Electromagnetic duality}
\label{dua}

The free Maxwell equations
\[
\vec{\nabla} \cdot \vec{E} = 0 \hspace{2cm} \vec{\nabla} \cdot \vec{B} = 0
\]
\beq
\vec{\nabla} \wedge \vec{E} + \frac{\partial \vec{B}}{\partial t}=0
\hspace{1cm}
\vec{\nabla} \wedge \vec{B} - \frac{\partial \vec{E}}{\partial t}=0
\label{Maxwell}
\eeq
are not only invariant under Lorentz and conformal transformations. They
are also invariant under a duality transformation :
\[
\vec{E} \rightarrow \cos \phi \vec{E} - \sin \phi \vec{B} 
\]
\beq
\label{duality}
\vec{B} \rightarrow \cos \phi \vec{B} + \sin \phi \vec{E} 
\eeq
In particular if we take $\phi = - \pi/2$ one obtains from eq. (\ref{duality})
a discrete duality transformation:
\beq
\vec{E} \rightarrow \vec{B} \hspace{2cm} \vec{B} \rightarrow - \vec{E}
\label{discdua}
\eeq
that is generated by the duality matrix  acting on the two-vector consisting
of the electric and magnetic fields as
\beq
\left( \begin{array}{c} \vec{E} \\ \vec{B} \end{array}  \right) 
\rightarrow 
\left( \begin{array}{cc} 0 & 1 \\ -1 & 0  \end{array}  \right)
\left( \begin{array}{c} \vec{E} \\ \vec{B} \end{array}  \right)
\label{discdua2}
\eeq

In terms of the  complex vector $\vec{E} + i \vec{B}$ the duality 
transformation in eq. (\ref{duality}) becomes
\beq
\vec{E} + i \vec{B} \rightarrow e^{i \phi} \left( \vec{E} + i \vec{B} \right)
\label{duacomple}
\eeq

Notice that the energy and momentum density of the electromagnetic field
given respectively by
\beq
\frac{1}{2} | \vec{E} + i \vec{B} |^2 = \frac{1}{2} \left(\vec{E}^2 + \vec{B}^2 
\right)
\label{energyden}
\eeq
and
\beq
\frac{1}{2i} (\vec{E} + i \vec{B})^{*} \wedge (\vec{E} + i \vec{B}) =
\vec{E} \wedge \vec{B}
\label{momdens}
\eeq
are invariant under the duality transformation in eq. (\ref{duality}), while 
the real and imaginary part of 
\beq
\frac{1}{2} \left( \vec{E} + i \vec{B} \right)^2 = \frac{1}{2} \left( 
\vec{E}^2 - \vec{B}^2 \right) + i \vec{E} \cdot \vec{B}
\label{dou}
\eeq
that are respectively the Lagrangian of the electromagnetic field and the 
topological charge density, transform as a doublet under the duality group.

If we perform a discrete duality transformations twice, 
we get
\beq
( \vec{E} , \vec{B} ) \rightarrow ( - \vec{E} , - \vec{B} ) 
\label{chconj}
\eeq
that corresponds to the charge conjugation operation.

The reason why this beautiful duality property of the free electromagnetic field
is not even mentioned in the courses on electromagnetism is due to the fact 
that it is lost when we introduce the interaction of the electromagnetic field
with matter by just adding in the right hand side of the Maxwell equations an
electric current $\vec{j}_e$ and an electric charge density $\rho_e$.
If we want to keep duality we must also introduce a magnetic current 
$\vec{j}_{m}$ and a magnetic charge density $\rho_{m}$ together with their 
electric counterparts. If we do so the Maxwell equations given in 
eq. (\ref{Maxwell}) and written in complex notations become:
\beq
\vec{\nabla} \cdot ( \vec{E} + i \vec{B} ) = \rho_e + i \rho_m
\label{Maxcomple1}
\eeq
and
\beq
\vec{\nabla} \wedge ( \vec{E} + i \vec{B} ) = i \frac{\partial}{\partial t}
(\vec{E} + i \vec{B} ) + j_e + i j_m
\label{Maxcomple2}
\eeq

The previous equations are invariant under the duality transformation given in
eq. (\ref{duacomple}) if the electric and magnetic currents and densities 
transform as
\beq
\rho_e + i \rho_m \rightarrow e^{i \phi} ( \rho_e + i \rho_m )
\label{duacomple1}
\eeq
and
\beq
j_e + i j_m \rightarrow e^{i \phi} ( j_e + i j_m )
\label{duacomple2}
\eeq
 
In particular if we have only pointlike particles with both electric and 
magnetic charge, then duality implies the following transformation:
\beq
q+ ig \rightarrow e^{i \phi} ( q +i g )
\label{duacomple3}
\eeq

Particles with magnetic charge are not  introduced in usual electromagnetism
for the very simple reason that they are not observed in the experiments. If
we include them we must either think that their mass is higher than the 
presently available energy or find other reasons for their absence. However,
if we insist in preserving duality also in the presence of interaction, as 
shown by Dirac~\cite{DIRAC}, a theory with both electric and magnetic charges
$q_i$ and $g_j$ can be consistently quantized only if  the Dirac
quantization condition  is satisfied 
\beq
q_i g_j = 2 \pi \, \hbar \, n_{ij}
\label{dirqua}
\eeq
where $n_{ij}$ are arbitrary integers. A complete discussion of the Dirac 
quantization condition can be found in the beautiful review by Goddard and
Olive~\cite{GO}.

The Dirac quantization condition is clearly not invariant under the duality 
transformation in eq. (\ref{duacomple3}). It is only invariant under the 
discrete transformation obtained from eq. (\ref{duacomple3}) for $\phi = - 
\frac{\pi}{2}$:
\beq
q \rightarrow g  \hspace{2cm} g \rightarrow - q
\label{disdua}
\eeq

\section{The 't Hooft-Polyakov monopole}
\label{mono}

In this section we will discuss the monopole solution found by {}'t 
Hooft~\cite{thooft} and Polyakov~\cite{Pol} in the Georgi-Glashow model. 
Many details are here omitted. The reader 
interested in them is recommended to consult Ref.~\cite{GO}.
 
Let us consider the Georgi-Glashow model
\beq
L = - \frac{1}{4} F_{a}^{\mu \nu} F_{a \,\, \mu \nu} + \frac{1}{2} (D_{\mu} 
\Phi )_a ( D^{\mu} \Phi )_a - V ( \Phi )
\label{ggmod}
\eeq
where
\beq
(D^{\mu} \Phi )_a  = \partial^{\mu} \Phi_a - e \epsilon_{abc} A^{\mu}_b \Phi_c
\label{covderi}
\eeq
and
\beq
F_{a}^{\mu \nu} = \partial^{\mu} A_{a}^{\nu} - \partial^{\nu} A_{a}^{\mu} -
e \epsilon_{abc} A_{b}^{\mu} A_{c}^{\nu}
\label{filstre}
\eeq
The gauge group is $SU(2)$ and the potential $V$ is equal to
\beq
V ( \Phi ) = \frac{\lambda}{4} \left( \Phi^2 - a^2 \right)^2
\label{pot}
\eeq
 
The classical equations of motion, that follow from $L$, are 
\beq
\left( D_{\nu} F^{\mu \nu} \right)_a = - e \epsilon_{abc} \Phi_b \left( D^{\mu}
\Phi \right)_c
\label{equ1}
\eeq

\beq
\left( D_{\mu} D^{\mu} \Phi \right)_a = - \lambda \Phi_a ( \Phi^2 - a^2 )
\label{equ2}
\eeq
They must be considered together with the Bianchi identity

\beq
D_{\mu} {}^* F^{\mu \nu} =0  \hspace{2cm} {}^* F^{\mu \nu} \equiv \frac{1}{2}
\epsilon^{\mu \nu \rho \sigma} F_{\rho \sigma}
\label{BIA}
\eeq
$\epsilon^{\mu \nu \rho \sigma}$ is the antisymmetric Levi Civita tensor 
with $\epsilon^{0123} =1$.

The energy is given by
\beq
E \equiv \int d^3 x \,\, \theta_{00} = \int d^3 x \left \{ \frac{1}{2} \left[ 
\left( B^{i}_{a} \right)^2  +  \left(  E^{i}_{a} \right)^2  + \Pi_{a}^{2}
+  \left( D^{i} \Phi \right)_{a}^{2} \right] + V( \Phi ) 
\right\} 
\label{ene}
\eeq
where
\beq
\Pi_a = \left(  D^0 \Phi \right)_a \hspace{1cm} F_{a}^{i0} = E_{a}^{i}
\hspace{1cm} F_{a \,\, ij} = - \epsilon_{ijk} B_{a}^{k}
\label{em}
\eeq

The energy is positive semi-definite. It vanishes if and only if
\beq
F^{\mu \nu}_a = \left( D^{\mu} \Phi \right)_a = 0 \hspace{1.5cm} V(\Phi ) =0
\label{vac}
\eeq
These conditions are satisfied by taking
\beq
\Phi_a = a \delta_{a3}  \hspace{2cm} A_{a}^{\mu} =0
\label{vac2}
\eeq
or equivalently any gauge rotated version of them.

This field configuration corresponds to the vacuum of our model and obviously
satisfies the equations of motions and the Bianchi identity in eqs. 
(\ref{equ1},\ref{equ2},\ref{BIA}).

It is easy to see that, if $ a \neq 0$, the $SU(2)$ gauge group is broken to 
$U(1)$. With the v.e.v of
the Higgs field taken along the third direction ($\Phi_a = a \delta_{a3}$) 
the $U(1)$  gauge field $A^{\mu}_{3}$ remains massless, while the two charged
fields
\beq
W_{\pm} = \frac{1}{\sqrt{2}}\left( A^{\mu}_{1} \pm i A^{\mu}_{2}
\right)
\label{www}
\eeq
get a mass equal to
\beq
M_{\pm} = a |q| 
\label{mass}
\eeq
where $q$ is their electric charge.
They are charged with respect to the unbroken $U(1)$ corresponding to the 
generator of the gauge group $SU(2)$ that leaves invariant the v.e.v of
the scalar field
\beq
Q = \frac{e}{a} T_a \Phi_a \hbar =  e T_3 \hbar 
\label{cha}
\eeq

Finally from the Higgs mechanism one gets also a neutral Higgs scalar particle
with mass equal to
\beq
M_H = \sqrt{2\lambda} \,\, a \,  \hbar
\label{Higmass}
\eeq
$\hbar$ has been explicitly written in some of the previous formulas,
while has been put equal to $1$ in most cases.

In addition to the constant vacuum solution of eq. (\ref{vac2})
the equations of motion admit also static (time independent) solutions. The
simplest of them can be obtained starting with a radially symmetric ansatz:
\beq
\Phi_a = \frac{r^a}{e r^2} H(aer) \hspace{1cm} A^{0}_{a} =0 \hspace{1cm}
A^{i}_{a} = - \epsilon_{aij}\frac{r^j}{e r^2} \left[1 - K (aer) \right]  
\label{ans}
\eeq

Inserting this ansatz into the energy one gets:
\beqa
E  & = & \frac{4 \pi a}{e} \int_{0}^{\infty} \frac{d\xi}{\xi^2} \left[ \xi^2 
\left( \frac{dK}{d \xi} \right)^2  + K^2 H^2 + \right. \nl  
& + & \left. \frac{1}{2} \left( \xi \frac{dH}{d \xi}
-H \right)^2 + \frac{1}{2} \left( K^2 -1 \right)^2 +  
\frac{\lambda}{4 e^2} \left(H^2 - \xi^2 \right)^2  \right]
\label{ene2}
\eeqa
where $\xi \equiv aer $ is a dimensionless quantity. The insertion of the 
ansatz in the equations of 
motions (\ref{equ1},\ref{equ2}) gives a system of coupled differential equations
for the radial functions $H$ and $K$:
\beq
\xi^2 \frac{d^2 K}{d \xi^2} = KH^2 + K (K^2 -1)
\label{equ3}
\eeq
and
\beq
\xi^2 \frac{d^2 H}{d \xi^2} = 2 K^2 H + \frac{\lambda}{e^2} H ( H^2 - \xi^2 )
\label{equ4}
\eeq

In order to have a finite energy solution one must also impose
boundary conditions for both  $\xi=0$ and $\xi \rightarrow \infty$. It can be 
shown that the previous system of equation admits a finite energy solution.
However, in general, it is not possible to write it down explicitly 
unless one takes the parameter $\lambda$ of the potential of the Higgs field  
equal to $0$. This corresponds to the so called Bogomolny~\cite{BOG}, Prasad, 
Sommerfield~\cite{PS} (BPS) limit. In this limit one obtains:
\beq
K ( \xi ) = \frac{ \xi }{\sinh \xi} \hspace{2cm} 
H( \xi) = \frac{ \xi }{\tanh \xi } -1
\label{bpslim}
\eeq

In order to have a better understanding of this limiting case let us rewrite
the sum of the two terms appearing in the energy density that involve
the square of the non abelian magnetic field and the square of the space 
components of the covariant derivative of the Higgs field as follows
\beq
\left( B^{i}_{a} \right)^2  +  \left( D^{i} \Phi \right)_{a}^{2} =
\left[   B^{i}_{a} \pm \left( D^{i} \Phi \right)_{a} \right]^2 \mp
2  B^{i}_{a} \left( D^{i} \Phi \right)_{a}
\label{ide}
\eeq

When we insert it in the energy (see eq. (\ref{ene})) we see that all terms
are positive except the last one in the r.h.s of eq. (\ref{ide}). We get 
therefore  a lower bound for the energy
\beq
E \geq \mp \int d^3 x   B^{i}_{a} \left( D^{i} \Phi \right)_{a}
\label{ine}
\eeq
that, after a partial integration and the use of the Bianchi identity in 
eq. (\ref{BIA}), becomes
\beq
E \geq \mp \int d^3 x \partial^i \left[ B^{i}_{a} \Phi_a \right]
\label{ine2}
\eeq

The equality sign is obtained if and only if the following equations are 
satisfied:
\beq
E_{a}^{i} = 0 \hspace{1cm}  \Pi_a = 0 \hspace{1cm}  \lambda =0
\hspace{1cm} B^{i}_{a} \pm \left( D^{i} \Phi \right)_{a} =0
\label{linequ}
\eeq
They are first order equations that imply the validity of the second order
equations of motion (\ref{equ1},\ref{equ2}). It can be checked that, if we
insert the ansatz in eq. (\ref{ans}) in the first order equations 
(\ref{linequ}) one obtains the solution  in eq. (\ref{bpslim}).

Inserting the static classical solution into the energy density given by the 
integrand in the r.h.s of eq. (\ref{ene}), one can see that it is concentrated
in a small region around the origin and goes to zero exponentially as $r$ goes
to infinity. 

In the quantum theory the classical solution corresponds to a new
particle of the spectrum that is an extended object (with size $\sim$ $1/a$) 
located in the region where
the energy density is appreciably different from zero. In this way we see that
the Georgi-Glashow model does not contain only perturbative states as a photon,
a massless Higgs field in the BPS limit
and a couple of charged bosons, all corresponding to the fields present in the 
Lagrangian in eq. (\ref{ggmod}) and having either a zero mass or a mass 
proportional to the gauge
coupling constant. It contains also additional particles that are soliton 
solutions of the classical equations of motion whose mass is instead 
proportional to the inverse of the gauge coupling constant as follows from eq.
(\ref{ene2}) and therefore are very massive in the weak coupling limit (small
$e$).

We want now to show that the soliton solution given by the ansatz in 
eq. (\ref{ans}) ia actually a magnetic monopole with respect to the unbroken
$U(1)$ group. 

It can be seen that for large enough values of $r$ the following equations are
satisfied:
\beq
D_{\mu} \Phi =0 \hspace{3cm}  \Phi^2 = a^2
\label{infi}
\eeq
apart from a small exponential correction.

Corrigan et al.~\cite{Corri} have shown that the most general solution of the 
previous equations corresponds to a vector field given by:
\beq
A^{\mu}_{a} = \frac{1}{a^2 e} \epsilon_{abc} \Phi_{b} \partial^{\mu} \Phi_{c}
+ \frac{1}{a} \Phi_{a} B^{\mu}
\label{solu}
\eeq
where $B^{\mu}$ is  arbitrary. The corresponding field strenght is equal to
\beq
F_{a}^{\mu \nu} = \frac{1}{a} \Phi_{a} F^{\mu \nu}
\label{fiestre}
\eeq
where
\beq
F^{\mu \nu} = \frac{1}{e a^3} \epsilon_{abc} \Phi_a \partial^{\mu} \Phi_b
\partial^{\nu} \Phi_c + \partial^{\mu} B^{\nu} - \partial^{\nu} B^{\mu}
\label{fiestre2}
\eeq

It satisfies the free Maxwell equations
\beq
\partial_{\mu} F^{\mu \nu} =0 \hspace{2cm}
\partial_{\mu} {}^{*}F^{\mu \nu} =0
\label{Maxeq}
\eeq
if the eqs. of motion in eqs. (\ref{equ1}), (\ref{equ2}) and (\ref{BIA}) are
satisfied.

We see that outside the region where the extended particle is located
the non abelian field strenght is aligned along the direction of the Higgs
field $\Phi_a $ and is proportional to an abelian field strenght 
$F^{\mu \nu}$ that can be interpreted as the field strenght of the unbroken 
$U(1)$ electromagnetic.

From eq. (\ref{fiestre}) we can compute the non abelian magnetic field that 
is equal to
\beq
B^{i}_{a} = - \frac{1}{2} \epsilon_{ijk} F_{a \,\, jk} = - \frac{1}{2 a^4 e } 
\Phi_{a} \epsilon_{ijk} \epsilon^{bcd} \Phi^b
\partial^{j} \Phi^{c} \partial^{k} \Phi^{d}
\label{nabemagn}
\eeq
Inserting it in eq. (\ref{ine2}) one gets:
\beq
E \geq \mp \frac{ 4 \pi a}{e} T 
\label{ine3}
\eeq
where the topological charge $T$ is given by   
\beq
T = \int d^3 x \;\; K_0 
\label{topcha}
\eeq
with
\beq
K_{\mu} = \frac{1}{8 \pi a^3} \epsilon_{\mu \nu \rho \sigma} \epsilon^{abc}
\partial^{\nu} \Phi^a \partial^{\rho} \Phi^b \partial^{\sigma} \Phi^c
\label{topcurr}
\eeq
We call it topological current because, unlike a Noether current, it is
conserved independently from the equations of motion as it can be trivially 
checked. It can also be seen (see Ref.~\cite{GO}) that the topological charge
$T$ is an integer since it counts the times that the two-sphere, defined by 
the second equation in eq. (\ref{infi}), is covered when the 
two-sphere at infinity in space is covered once.

In conclusion we get
\beq
E \geq \mp a g 
\label{enemagn}
\eeq
where $g$ is the magnetic charge of the soliton solution that is 
obtained by integrating the equation:
\beq
\partial^i B^i =  \frac{4 \pi}{e} K_0
\label{divb}
\eeq
that follows from eq. (\ref{fiestre2}). One gets:
\beq
g = \int d^3 x \partial^i B^i =  \frac{4\pi}{e} T
\label{magncha}
\eeq

In the case of the static solution corresponding to the ansatz in eq. 
(\ref{ans}) it is easy to see that $T = \mp 1$ in such a way that $E \geq
a |g| $.

The value obtained for the magnetic charge $g = \mp \frac{4 \pi}{e}$ is
consistent with the Dirac quantization condition (with n=2) given in eq.
(\ref{dirqua})
\beq
q g = 4 \pi \hbar
\label{dirac2}
\eeq
where $q$ is the charge of the $W$-boson given in eq. (\ref{cha}) for $T_3 
= \pm 1$. The fact 
that we obtain $n=2$ is a consequence of the fact that the gauge bosons 
transform according to the triplet representation of the gauge group $SU(2)$.
The value $n=1$ would have been obtained with matter fields transforming
according to the fundamental doublet representation.

The soliton solution following from the ansatz in eq. (\ref{ans}) has a 
non vanishing magnetic charge, but has zero electric charge. In order to also 
have a solution with a non vanishing electric charge~\cite{juzee} we must 
allow for a non vanishing electric potential of the type
\beq
A^{0}_{a} = \frac{r^a}{e r^2} J (r)
\label{dyonansa}
\eeq
instead of the vanishing ansatz given in eq. (\ref{ans}). With this ansatz 
the equations of 
motion in eqs. (\ref{equ3},\ref{equ4}) are modified as follows:
\beq
\xi^2 \frac{d^2 K}{d \xi^2} = K \left[ K^2 + H^2 - J^2 -1 \right]
\label{equ3'}
\eeq
and
\beq
\xi^2 \frac{d^2 H}{d \xi^2} = 2 K^2 H + \frac{\lambda}{e^2} H ( H^2 - \xi^2 )
\hspace{1cm}  \xi^2 \frac{d^2 J}{d \xi^2} = 2 K^2 J
\label{equ4'}
\eeq
In the BPS limit where $\lambda=0$ one can obtain an analytical solution
given by:
\beq 
H ( \xi ) = \cosh \gamma \left[ \frac{\xi}{\tanh \xi} -1 \right]
\hspace{1cm}
K (\xi ) = \frac{\xi}{\sinh \xi} 
\label{dyonsol1}
\eeq
and
\beq
J ( \xi ) = \sinh \gamma \left[ \frac{\xi}{\tanh \xi} -1 \right]
\label{dyonsol2}
\eeq
where $\gamma$ is an arbitrary constant.

This solution corresponds to an extended object
that has both electric and magnetic charge given  by
\beq
q \equiv \int d^3 x \partial^{i} E^{i} = \frac{4 \pi}{e} \sinh \gamma
\hspace{2cm} g =  \frac{4 \pi}{e} T
\label{magele}
\eeq
while its mass is given by:
\beq
M = \frac{4 \pi}{e} a \cosh^{2} \gamma
\label{dyonmass}
\eeq
Using the asymptotic behaviour for large $r$ of the Higgs field
\beq
\Phi^a = \frac{r^a}{r} \frac{H(ear)}{ er} \rightarrow \frac{r^a}{r} v
\label{asympt}
\eeq
where $v \equiv a \cosh \gamma$, together with the expressions for the electric
and magnetic charges given in eq. (\ref{magele}) we obtain the mass of the dyon:
\beq
M = v \sqrt{q^2 + g^2}
\label{dyonmass2}
\eeq
This formula has been deduced for the dyon soliton solution in the BPS limit, 
but  it is actually valid for any 
particle of the spectrum. Notice that it is invariant under the duality 
transformation in eq. (\ref{duacomple3}).

In the classical theory the electric charge of the dyon can get any value 
given by the formula
in eq. (\ref{magele}). When one semiclassically quantizes the dyon solution one
discovers that
the values for the electric charges are quantized~\cite{OSBORN2} and given by
\beq
q= e n_e 
\label{quaelecha}
\eeq
where $n_e$ is an integer.

\section{Montonen-Olive duality}
\label{MOdua}

Leaving aside for a moment the dyon solution discussed at the end of last 
section we have 
found  that the semiclassical spectrum of the Georgi-Glashow model in the 
BPS limit consists of a massless photon and Higgs particle, of an electrically
charged $W$ boson with charge equal to $ q_0 = \pm e \hbar$ and of a 
magnetic monopole with magnetic charge equal to $g_0 = \pm \frac{4 \pi}{e} =
\frac{4 \pi \hbar}{q_0}$.

If there is duality invariance as suggested for instance by the formula in eq. 
(\ref{dyonmass2}) we can make a duality transformation
with angle $\phi = - \frac{\pi}{2}$ such that
\beq
q_0 \rightarrow g_0 \hspace{2cm} g_0 \rightarrow - q_0
\label{duality2}
\eeq
This transformation implies that
\beq
q_0 \rightarrow \frac{4 \pi \hbar }{q_0}
\label{duality3}
\eeq

Based on this observation Montonen and Olive~\cite{MONTOLI} suggested that 
there are
two equivalent formulations of the same theory dual to each other. In the 
first one, that we call electric, the $W$'s are elementary particles while
the magnetic monopoles are solitons. In the second one, that we call magnetic,
the elementary particles are instead the magnetic monopoles while the $W$ 
bosons are solitons. They
also suggested  that the two formulations had essentially the same Lagrangian.
The only important difference between them is that the
electric theory is weakly coupled when $q_0 \rightarrow 0$ ($e \rightarrow 0$)
while the magnetic theory is weakly coupled when $g_0 \rightarrow 0$ 
corresponding to $e \rightarrow \infty$. They brought the following arguments
in support of their duality conjecture.
 
\begin{enumerate}
\item{The mass formula in eq. (\ref{dyonmass2}), valid for all particles of the 
theory, is duality invariant.}
\item{Since there is no interaction between two monopoles, while there is a
non zero interaction between a monopole and an antimonopole, if duality is
correct, one must expect that the interaction between equal charge $W$-bosons
must be zero while that between  opposite charged $W$-bosons must be non 
vanishing.
This is actually verified in the BPS limit because in this limit the Higgs 
field is also
massless and  contributes with opposite sign with respect to the photon for 
equal charge 
$W$, while it contributes with opposite  sign for opposite charge $W$.}
\end{enumerate}

The Montonen-Olive duality proposal, leaves, however many unanswered 
questions that we list:

\begin{enumerate}
\item{The elementary $W^{\pm}$ bosons have spin equal to $1$. If the 
magnetic monopoles are dual to them they must also have spin equal to $1$. 
But how can this happen?}

\item{The previous considerations are based on a mass formula that is only 
valid classically. How are the quantum corrections going to modify it?}

\item{In the previous considerations we have neglected the dyons. What is their 
role in the all picture?}
\end{enumerate}

The previous questions do not have an answer in the framework of the 
Georgi-Glashow model discussed in the previous section since the quantum 
Georgi-Glashow model is, actually, not duality invariant. But it was soon 
recognized~\cite{DADDA} that, in order to have a theory with Montonen-Olive 
duality, one
must include supersymmetry since in a supersymmetric theory the quantum 
corrections coming from the bosons and the fermions tend to cancel each others
preserving the structure of the classical mass formula. Actually the argument 
used  in Ref.~\cite{DADDA} for the $N=2$ super Yang-Mills theory is too naive 
and in fact wrong
as pointed out in Refs.~\cite{SCHONFELD,KAUL83,KAUL84,IMBIMBO84,IMBIMBO85} 
because this theory is not ultraviolet
finite. In order to have a classical mass formula that is not modified by 
quantum corrections one must consider the $N=4$ super Yang-Mills theory that 
is free from ultraviolet divergences~\cite{MANDE,BRINK}, as it was done by 
Osborn~\cite{OSBORN} who made also the important observation that in this case
magnetic monopoles and dyons have also supersymmetric partners with spin equal
to $1$. The introduction of the $N=4$ theory open the way to the solution of
the first two puzzles discussed above. In the meantime Witten and 
Olive~\cite{WO} found
out that the structure of the duality invariant mass formula in eq. 
(\ref{dyonmass2}) for a BPS state in the $N=2$ theory is a direct consequence
of the supersymmetry algebra opening the way to the quantum exact determination
of the mass of the BPS states. This observation is  playing an essential
role also in recent developments in string theories.

\section{Representations of supersymmetry algebra}
\label{repre}

As in the case of the Poincar{\'{e}} group the representations of the 
supersymmetry algebra for massive particles are different  from those
for massless particles. The supersymmetry algebra is given in both cases by:
\beq
\{ Q_{\alpha}^{i} , {\bar{Q}}_{\dot{\alpha}}^{j} \} = 2 \sigma_{\mu} P^{\mu} 
\delta^{ij}  \hspace{2cm} i,j=1 \dots N
\label{sualg}
\eeq
\beq
\{ Q_{\alpha}^{i} , Q_{\beta}^{j} \} =
\{ {\bar{Q}}_{\dot{\alpha}}^{i} , {\bar{Q}}_{\dot{\beta}}^{j} \} = 0
\label{sualg2}
\eeq
The difference between the two cases is due to the fact that in the massive 
case one can
always choose a center of mass frame where $P_{\mu} = ( M , \vec{0} )$, while 
this is not possible in the massless case. 

In the massive case in the center of mass frame one gets the following algebra:
\beq
\{ a^{i}_{\alpha} , (a^{j}_{\beta})^{+} \} = \delta_{\alpha \beta} \delta^{ij}
\label{sualg3}
\eeq
and
\beq
\{ a^{i}_{\alpha} , a^{j}_{\beta} \} = 
\{ (a^{i}_{\alpha})^{+}  , (a^{j}_{\beta})^{+} \} = 0
\label{sualg4}
\eeq
where
\beq 
 a^{i}_{\alpha}  = \frac{1}{\sqrt{2 M}} Q_{\alpha}^{i} \hspace{1cm}
 (a^{j}_{\beta})^{+}  = \frac{1}{\sqrt{2 M}} {\bar{Q}}^{i}_{\dot{\alpha}}
\label{defi1}
\eeq

The representation of the fermionic harmonic oscillator algebra is constructed
starting from a vacuum state $|0>$ satisfying the equation:
\beq
a_{\alpha}^{i} | 0 > =0
\label{vac1}
\eeq
and acting on it with the creation operators:
\beq
\frac{1}{\sqrt{n!}} ( a_{\alpha_1}^{i_1} )^{+} (a_{\alpha_2}^{i_2} )^{+} 
\dots (a_{\alpha_n}^{i_n} )^{+} |0 > \hspace{1cm} n = 0,1 \dots 2N 
\label{state1}
\eeq
The number of states  in eq. (\ref{state1}) is equal to
$ \left( \begin{array}{c} 2N \\
                           n  \end{array} \right)$.
Since $n$ runs from $0$ to $2N$
the total number of states in the representation of the massive supersymmetry
algebra is equal to:
\beq
\sum_{n=0}^{2N} \left( \begin{array}{c} 2N \\
                           n  \end{array} \right) = 2^{2N}
\label{totsta}
\eeq
The states in the representation have a maximum helicity gap $\Delta \lambda =
N$. Half of them are fermions and the other half are bosons.

In the massless case we can instead choose a frame where $P_{\mu} = (E,0,0,-E)$.
In this frame the supersymmetry algebra becomes:
\beq
\{ a^i , ( a^{j} )^{+} \} = \delta^{ij}
\label{zerosal}
\eeq
\beq
\{ a^{i} , a^{j} \} = 
\{ (a^{i})^{+}  , (a^{j})^{+} \} = 0
\label{sualg5}
\eeq
where  
\beq 
 a^{i}  = \frac{1}{2 \sqrt{E}} Q_{1}^{i} \hspace{1cm}
 (a^{j})^{+}  = \frac{1}{2 \sqrt{E}} {\bar{Q}}^{i}_{\dot{1}}
\label{defi2}
\eeq
The anticommutators involving the generators of the supersymmetry algebra with
indices $ \alpha =2$ and $ {\dot{\alpha}} = \dot{2}$ are all vanishing and
therefore they can be consistently put equal to zero:
\beq
Q^{i}_{2} = {\bar{Q}}^{i}_{\dot{2}} =0
\label{zecomm}
\eeq

Starting again from the vacuum state annihilated by the annihilation operators
$a^{i}$ we can construct the states of the representation acting on it
with the creation operators obtaining the state:
\beq
\frac{1}{\sqrt{n!}} ( a^{i_1} )^{+} ( a^{i_2} )^{+} 
\dots (a^{i_n} )^{+} |0 > \hspace{2cm} n=0,1 \dots N 
\label{state2}
\eeq
that contains $ \left( \begin{array}{c} N \\
                                        n \end{array} \right)$ states. 
The total number of states in the massless representation is equal to:
\beq
\sum_{n=1}^{N}  \left( \begin{array}{c} N \\
                                        n \end{array} \right) = 2^{N}
\label{totsta2}
\eeq
that is smaller than in the case of a massive representation. The maximum 
helicity in this case is $ \Delta \lambda = N/2$. In the case $N=1$ one gets only 
one fermionic and one bosonic
state. In most cases, however, we must add another multiplet with opposite
helicity in order to have a $CPT$ invariant theory ($CPT$ reverses the sign of
helicity).

Let us finally consider the representation of the massive $N=2$ algebra with
non vanishing central charges~\cite{HAAG}. In this case in the center of mass 
frame the algebra is 
\beq
\{ Q_{\alpha}^{i} , {\bar{Q}}_{\dot{\alpha}}^{j} \} = 2 M 
\delta^{ij}  \delta_{\alpha {\dot{\alpha}}} 
\label{sualgcc}
\eeq
\beq
\{ Q_{\alpha}^{i} , Q_{\beta}^{j} \} = \epsilon^{ij} \epsilon_{\alpha \beta}
Z \hspace{2cm}
\{ {\bar{Q}}_{\dot{\alpha}}^{i} , {\bar{Q}}_{\dot{\beta}}^{j} \} = 
\epsilon^{ij} \epsilon_{\dot{\alpha}  \dot{\beta}} \bar{Z}
\label{sualgcc2}
\eeq
One can get rid of the phase in $Z$ by a supercharge redefinition. Then one 
can rewrite the previous algebra in terms of the two quantities:
\beq
a_{\alpha} = \frac{1}{\sqrt{2}} \left[ Q^{1}_{\alpha} + \epsilon_{\alpha \beta}
{\bar{Q}}^{2}_{\beta} \right] \hspace{2cm}
b_{\alpha} = \frac{1}{\sqrt{2}} \left[ Q^{1}_{\alpha} - \epsilon_{\alpha \beta}
{\bar{Q}}^{2}_{\beta} \right] 
\label{defnew}
\eeq
obtaining
\beq
\{ a_{\alpha} , (a_{\beta} )^{+} \} = ( 2 M + |Z| ) \delta_{\alpha \beta}
\hspace{1cm}
\{ b_{\alpha} , (b_{\beta} )^{+} \} = ( 2 M - |Z| ) \delta_{\alpha \beta}
\label{defnew2}
\eeq
while all the other anticommutators are vanishing.

If $2M = |Z|$ all anticommutators involving the oscillators $b$ are vanishing
and therefore we can put them equal to zero. We can then use only the 
oscillators $a$ for constructing the representation, obtaining the same number
of states as in the massless case. In the case $N=2$ here considered we get
the following four states:
\beq
|0>  \hspace{2cm} a^{+}_{\alpha} | 0> \hspace{2cm} a^{+}_{\alpha} 
a^{+}_{\beta} | 0> 
\label{staneq2}
\eeq
instead of the $16$ states that we found in the case without central charge
(see eq. (\ref{totsta}) for $N=2$). 

Extending the previous procedure to the case $N=4$ we obtain a short 
representation 
with $16$ states instead of the one with $2^8 = 256$ states obtained without 
central charge (See eq. (\ref{totsta}) for $N=4$).

The fact that the representations of extended supersymmetry with non vanishing 
central charges are shorter and have the same dimension of those for the 
massless case makes it possible to have a consistent supersymmetric Higgs 
mechanism since in this case one has the
same number of degrees of freedom before and after the Higgs mechanism.
 
\section{Supersymmetric actions}
\label{supe}

In this section we construct the supersymmetric extension of Yang-Mills theory
in $D=4$ by dimensional reduction from higher dimensions~\cite{BSS}.

Let me start from the following action in $D$ dimensions
\beq
S = \int d^D x \left\{ - \frac{1}{4} F_{M N}^{a} F^{a \, M N} - 
\frac{i}{2} {\bar{\lambda}}^a \Gamma_{M} (D^{M} \lambda )^a \right\}
\label{susyd}
\eeq

If we perform the following supersymmetry transformation
\beq
\delta A_{M}^{a} =  \frac{i}{2} \left[ {\bar{\lambda}}^a \Gamma_{M} \alpha
- {\bar{\alpha}} \Gamma_{M} \lambda^{a} \right]
\label{sutra1}
\eeq
together with
\beq
\delta \lambda_a = \sigma_{R S} F_{a}^{RS} \alpha
\hspace{2cm}
\delta {\bar{\lambda}}_a = - {\bar{\alpha}} \sigma_{RS} F_{a}^{RS} 
\label{sutra2}
\eeq
it can be seen, by using the useful identities,
\beq
\Gamma^{M} \sigma^{R S} = \frac{1}{2} \left[ g^{M R} \Gamma^{S} - 
g^{M S} \Gamma^{R} - \frac{1}{(D-3)!} \epsilon^{M R S N_1 \dots 
N_{D-3}} \Gamma_{D+1} \Gamma_{N_1} \dots \Gamma_{N_{D-3}}  
\right]
\label{ide1}
\eeq
and
\beq
\sigma^{R S} \Gamma^{M} = \frac{1}{2} \left[ - g^{M R} \Gamma^{S} + 
g^{M S} \Gamma^{R} - \frac{1}{(D-3)!} \epsilon^{M R S N_1 \dots 
N_{D-3}} \Gamma_{D+1} \Gamma_{N_1} \dots \Gamma_{N_{D-3}}  
\right]
\label{ide2}
\eeq
that the term with the $\epsilon$ tensors cancel using the Bianchi identity
for $F_{\mu \nu}$ and the action in eq. (\ref{susyd}) transforms as a total 
derivative 
\beq
\delta S = \frac{i}{2} \int d^{D} x \partial_{M} \left[  {\bar{\alpha}} 
\sigma_{R S}
F_{a}^{R S} \Gamma^{M} \lambda_{a} +  F_{a}^{M N} \left( {\bar{\alpha}} 
\Gamma_{N} \lambda^{a} - {\bar{\lambda}}^{a} \Gamma_{N} \alpha \right) 
\right]
\label{sustra}
\eeq
provided that the following equation is satisfied
\beq
\left({\bar{\lambda}}^{a} \Gamma_{M} f^{abc} \lambda^{c}  \right)  
\left[ {\bar{\alpha}} 
\Gamma^{M} \lambda^{b} - {\bar{\lambda}}^{b} \Gamma^{M} \alpha \right] =0
\label{equasu}
\eeq
$\alpha$ and $\lambda$ are spinors in $D$-dimensions, 
$\sigma_{M N} = \frac{1}{4} [ \Gamma_{M} , \Gamma_{N}] $, $\Gamma_{D+1} =
\Gamma_0 \dots \Gamma_{D}$ and 
\beq
F^{a}_{M N} = \partial_{M} A_{N}^{a} - \partial_{N} A_{M}^{a} - e f^{abc} 
A_{M}^{b}
A_{N}^{c} \hspace{2cm} 
(D_{M} \lambda )^a = \partial_{M} \lambda^{a} - e f^{abc} A_{M}^{b}
\lambda^{c}
\label{defi}
\eeq

Therefore the action in eq. (\ref{susyd}) is $N=1$ supersymmetric if the 
equation (\ref{equasu}) is satisfied. As shown in Ref.~\cite{BSS} this 
happens in the following cases:
\begin{enumerate}
\item{D=3, if $\lambda$ is a Majorana spinor }
\item{D=4, if $\lambda$ is a Majorana spinor }
\item{D=6, if $\lambda$ is a a Weyl spinor }
\item{D=10, if $\lambda$ is a Weyl-Majorana spinor}
\end{enumerate}
There is a simple way to understand this result by noticing that,
in all these cases, the number of on shell bosonic degrees of freedom, that is 
equal to $D-2$, is 
equal to the number of on shell fermionic degrees of freedom that is equal to 
$2^{[\frac{D}{2}]}$
multiplied with a factor $x =\frac{1}{2}$ if the spinor field $\lambda$ is a 
Majorana or Weyl spinor
and a factor $ x = \frac{1}{4}$ if the spinor field is a Weyl-Majorana spinor:
\beq
D-2 =  x \;\; 2^{[\frac{D}{2}]}
\label{degfre}
\eeq
where $[\frac{D}{2}]= \frac{D}{2}$ if $D$ is even and $[\frac{D}{2}] = 
\frac{D-1}{2}$ if $D$ is odd.

In particular 
the action in eq. (\ref{susyd}) is $N=1$ supersymmetric if $D = 6 $ and $
10$. This fact can be used to write actions with extended $N=2,4$ 
supersymmetries in
four dimensions by the technique of dimensional reduction. Let us divide the
$D$ dimensional space-time component $x^{M} \equiv ( x^{\mu} , x^{i} )$ in a 
part $x^{\mu}$, where the index $\mu$ runs over the four-dimensional 
space-time, and in part $x^{i}$, where the index $i$ runs over the compactified
$D-4$ dimensions. We assume that the various fields are independent from
the compactified coordinates. 

Let us start to compactify the bosonic term in the action (\ref{susyd})
containing the non abelian field  strenght given in eq. (\ref{defi}). The
dimensional reduction of $F_{MN}$ gives respectively:
\beq
F^{a}_{\mu \nu } = \partial_{\mu} A_{\nu}^{a} - \partial_{\nu} A_{\mu}^{a} - 
e f^{abc} A_{\mu}^{b} A_{\nu}^{c}
\label{fmunu}
\eeq
\beq
F^{a}_{\mu i} = \partial_{\mu} A^{a}_{i} - e f^{abc} A_{\mu}^{b} A_{i}^{c}
\equiv \left( D_{\mu} A_{i} \right)^a
\label{fmui}
\eeq
and
\beq
F_{ij}^{a} = - e f^{abc} A_{i}^{b} A_{j}^{c}
\label{fij}
\eeq
Using the previous equations one obtains immediately the compactification
of the gauge kinetic term
\beq
- \frac{1}{4} F_{MN}^a F^{a \,MN} = - \frac{1}{4} F^{a}_{\mu \nu} F^{a \,
\mu \nu}
 + \frac{1}{2} \left( D_{\mu} A_{i} \right)^{a} \left( D^{\mu} A_{i} 
\right)^{a} - \frac{e^2}{4} f^{abc} A_{i}^{b} A_{j}^{c}
f^{ade} A_{i}^{d} A_{j}^{e}
\label{combos}
\eeq
where a sum over repeated indices is understood.

In order to perform the compactification of the fermionic term of the action
in eq. (\ref{susyd}) we have to distinguish the two cases $D=6$ and $D=10$.

A representation of the Dirac algebra for $D=6$ is given by:
\beq
\Gamma_{\mu} = \gamma_{\mu} \otimes 1  \hspace{1cm} \mu = 0,1,2,3 
\label{gamma6}
\eeq
\beq
\Gamma_{4} = \gamma_5 \otimes i \sigma_{1}  \hspace{1cm}
\Gamma_{5} = \gamma_5 \otimes i \sigma_{2} \hspace{1cm} \Gamma_7 \equiv
\Gamma_0 \Gamma_1 \Gamma_2 \Gamma_3 \Gamma_4 \Gamma_5 = \gamma_5
\otimes \sigma_3 
\label{gamma6b}
\eeq
where the $\sigma$-matrices are the Pauli matrices and $\gamma_5 = 
i \gamma^{0} \gamma^{1} \gamma^{2} \gamma^{3}$. 

A Weyl spinor in $D=6$ satisfies the condition:
\beq
( 1 + \Gamma_7 ) \lambda =0
\label{weylco}
\eeq
that is automatically satisfied if we take
\beq
\lambda = \left( \begin{array}{c} \frac{1 - \gamma_5 }{2} \chi \\ 
                                  \frac{1 + \gamma_5 }{2} \chi
                    \end{array} \right)
\hspace{2cm}
{\bar{\lambda}} = \left( \begin{array}{cc} 
{\bar{\chi}} ( \frac{1+ \gamma_5}{2} ) & 
{\bar{\chi}} (\frac{1- \gamma_5}{2}) \end{array} \right)
\label{sol}
\eeq
where $\chi$ is a Dirac spinor in four dimensions.

Inserting it in the fermionic term in eq. (\ref{susyd}) one gets:
\beq
i \left({\bar{\lambda}}\right)^a \Gamma^{M} D_{M} \lambda^a = i 
\left( {\bar{\chi}} \right)^a \Gamma^{\mu} \left( D_{\mu}
\chi \right)^a - e f^{abc} {\bar{\chi}}^{a} A_{4}^{b} \gamma_5 \chi^{c} + ie 
f^{abc} {\bar{\chi}}^{a} A_{5}^{b} \chi^{c}   
\label{comfer}
\eeq

The Lagrangian of $N=2$ super Yang-Mills is obtained by summing the bosonic
contribution in eq. (\ref{combos}) with the indices $i,j=1,2$ to the
fermionic contribution in eq. (\ref{comfer}). One gets
\[
L= - \frac{1}{4} F^{a}_{\mu \nu} F^{a \, \mu \nu}
 + \frac{1}{2} \sum_{i=1}^{2} \left( D_{\mu} A_{i} \right)^{a} 
\left( D^{\mu} A_{i} 
\right)^{a} - \frac{e^2}{2} f^{abc} A_{4}^{b} A_{5}^{c}
f^{ade} A_{4}^{d} A_{5}^{e} +
\] 
\beq
- i \left( {\bar{\chi}} \right)^a \Gamma^{\mu} \left( D_{\mu}
\chi \right)^a  + e f^{abc} {\bar{\chi}}^{a} A_{4}^{b} \gamma_5 \chi^{c} - ie 
f^{abc} {\bar{\chi}}^{a} A_{5}^{b} \chi^{c} 
\label{neq2sYM}
\eeq
after a redefinition of the Dirac spinor $ \chi \rightarrow \sqrt{2} \chi$.

The $N=4$ super Yang-Mills is instead obtained starting with a Weyl-Majorana
spinor in $D=10$. In $D=10$ the Dirac algebra can be represented
as follows:
\beq
\Gamma^{\mu} = \gamma^{\mu} \otimes 1 \otimes \sigma_3 \hspace{1cm}
\mu=0,1,2,3
\label{gmu}
\eeq
\beq
\Gamma^{3+i} = 1 \otimes \alpha^i \otimes \sigma_1 \hspace{1cm}
\Gamma^{6+i} = \gamma_5 \otimes \beta^i \otimes \sigma_3 \hspace{1cm}
i=1,2,3
\label{gi}
\eeq
where  the fourdimensional internal 
matrices $\alpha$ and $\beta$ satisfy the following algebra:
\beq
\{\alpha^i , \alpha^j \} = \{\beta^i , \beta^j \} = -2 \delta^{ij} 
\hspace{1cm} [ \alpha^i , \beta^j ] =0
\label{con}
\eeq
and
\beq
[ \alpha^i , \alpha^j ] = -2 \epsilon^{ijk} \alpha^{k} \hspace{2cm}
[ \beta^i , \beta^j ] = -2 \epsilon^{ijk} \beta^{k}
\label{excon}
\eeq
Finally the correspondent of $\gamma_5$ in ten dimensions is given by:
\beq
\Gamma_{11} = \Gamma_0 \Gamma_1 \Gamma_2 \Gamma_3 \Gamma_4 \Gamma_5 \Gamma_6
\Gamma_7 \Gamma_8 \Gamma_9 = 1 \otimes 1 \otimes \sigma_2
\label{chiope}
\eeq

A Weyl-Majorana spinor satisfying the condition:
\beq
( 1 + \Gamma_{11} ) \lambda =0
\label{Weyco}
\eeq
can always be written as 
\beq
\lambda = \psi \otimes \frac{1}{ \sqrt{2}} \left( \begin{array}{c}
                                                        1 \\
                                                        -i
                                  \end{array} \right)
\hspace{1cm}
{\bar{\lambda}} = {\bar{\psi}} \frac{1}{\sqrt{2}} \left( \begin{array}{cc}
                                                    1 & -i 
                                                \end{array} \right)
\label{weyspi}
\eeq
where the Majorana spinor $\psi$ has a four dimensional space-time index 
on which the Dirac matrices act and another internal four dimensional index
on which instead the internal matrices $\alpha$ and $ \beta$ act. 

Proceeding as in the $N=2$ case we arrive at the $N=4$ super Yang-Mills
Lagrangian:
\[
L=  - \frac{1}{4} F^{a}_{\mu \nu} F^{a \, \mu \nu}
 + \frac{1}{2} \sum_{i=1}^{3} \left( D_{\mu} A_{i} \right)^{a} 
\left( D^{\mu} A_{i} 
\right)^{a} + \frac{1}{2} \sum_{i=1}^{3} \left( D_{\mu} B_{i} \right)^{a} 
\left( D^{\mu} B_{i} \right)^{a} - V(A_i , B_j) + 
\]
\beq
- \frac{i}{2} \left( {\bar{\psi}} \right)^a \gamma^{\mu} \left( D_{\mu} \psi 
\right)^a - \frac{e}{2}  
f^{abc} {\bar{\psi}}^{a} \alpha^{i} A^{b \, i} \psi^{c} - i 
\frac{e}{2} 
f^{abc} {\bar{\psi}}^{a} \beta^{j} \gamma_5 B^{b \, j} \psi^{c} 
\label{neq4sYM}
\eeq
where the potential is equal to:
\beq
V( A_i , B_j ) = \frac{e^2}{4} f^{abc} A_{i}^{b} A_{j}^{c}
f^{afg} A_{i}^{f} A_{j}^{g} + \frac{e^2}{4} f^{abc} B_{i}^{b} B_{j}^{c}
f^{afg} B_{i}^{f} B_{j}^{g} + \frac{e^2}{2} f^{abc} A_{i}^{b} B_{j}^{c}
f^{afg} A_{i}^{f} B_{j}^{g}
\label{potn=4}
\eeq

\section{Semiclassical analysis of super $N=2$ theory}
\label{semi}

The $N=2$ super Yang-Mills theory described by the Lagrangian in eq. 
(\ref{neq2sYM}) can be rewritten in the following form
\[
L = \frac{1}{4 \pi} Im \left\{ \left( 
\frac{\theta}{2 \pi} + i \frac{4 \pi}{e^2} \right) 
\left[ - \frac{1}{4}\left(
F^{a}_{\mu \nu}  F^{a}_{\mu \nu}  - \frac{1}{2} \epsilon^{\mu \nu \rho 
\sigma} F^{a}_{\mu \nu} F^{a}_{\rho \sigma} \right) + \right. \right. 
\]
\beq
\left. \left. +  \overline{(D_{\mu} \Phi)}^a  (D_{\mu} \Phi)^a -
\frac{1}{2} [ f^{abc} \Phi^b {\bar{\Phi}}^c ]^2 + \;\; FERMIONS
\right] \right\}
\label{newneq2}
\eeq
after a rescaling by a factor $1/e$ of the fields and the introduction of the 
vacuum angle $\theta$ and of a complex field $\Phi = 
\frac{A_5 + i A_4}{\sqrt{2}}$.

The structure of the bosonic part of the Lagrangian in eq. (\ref{newneq2}) is
pretty much the same as the one in the Georgi-Glashow model in eq. 
(\ref{ggmod}). There is, however, an important difference. Unlike the 
Georgi-Glashow here the potential given in the case of a $SU(2)$ gauge group
by
\beq
\label{pot8}
V ( \Phi ) = \frac{1}{2e^2} [ \epsilon^{abc} \Phi^b {\bar{\Phi}}^c ]^2 
\eeq
does not fix uniquely the vacuum. In fact any field configuration of the type
\beq
\label{pot9}
\Phi^a = ( 0, 0 , a )
\eeq
corresponds to a minimum of the potential with vanishing value (since 
supersymmetry is not broken) for any value of the complex variable $a$.
The set of all values of $a$ is called the classical moduli space of
the theory. Actually a better parametrization of the vacua is  given in terms 
of the gauge invariant variable $u = \frac{1}{2} a^2 = Tr (\Phi^2) $. 

If $a \neq 0$, as in the Georgi-Glashow model, the $SU(2)$ gauge symmetry is 
broken to $U(1)$ by the Higgs phenomenon and the charged (with respect to the
unbroken $U(1)$) components of the gauge fields $W^{\pm}$ get a non  vanishing
mass, while the Higgs and the gauge field of the unbroken $U(1)$ remain 
massless as in the Georgi-Glashow model in the BPS limit. In the supersymmetric
case, however, the BPS limit is obtained without needing to send to zero 
any piece of the potential as it was necessary in the Georgi-Glashow model.    

As in the Georgi-Glashow model, in the $N=2$ super Yang-Mills theory there are 
also time-independent solutions~\cite{DADDA} of the classical 
equations of motion corresponding to magnetic monopoles and dyons. 

Their mass can again be written in terms of the electric and magnetic charges
as in the Georgi-Glashow model
\beq
M= \sqrt{2}|a|\sqrt{g^2 + q^2}
\label{mass8}
\eeq
where $a$ has been defined in eq. (\ref{pot9}).

This formula for the mass is again valid for all particles of the semiclassical
spectrum.

After semiclassical quantization the electric and magnetic charges 
of the particles of the spectrum are given by
\beq
g= \frac{4 \pi}{e}n_m  \hspace{1cm} q= e n_e 
\label{elema}
\eeq
where $n_m = \pm 1$, $n_e$ is an integer and their mass is given 
by~\cite{KAUL84,IMBIMBO85}
\beq
\label{mass4}
M= \sqrt{2}|Z| \hspace{2cm}
Z= a \left[ n_e + \left( \frac{\theta}{2 \pi} + i \frac{4 \pi}{e^{2}(\mu) } 
+ \frac{i}{\pi} \log \frac{a^2}{\mu^2 C} \right) 
n_m \right]
\eeq
where $\mu$ is the renormalization scale and $C$ is a scheme dependent constant.
 
As a consequence of the supersymmetric Higgs mechanism the massless spectrum 
consists of a photon $A_{\mu}$, that is the gauge field of the unbroken $U(1)$,
of a photino and of a complex scalar particle $A$. They belong to a 
massless $N=2$ chiral supermultiplet. If we are interested in studying the 
low-energy
dynamics of these fields  and therefore we need to restrict ourselves to a
Lagrangian with at most two derivatives and with no more than four-fermion 
couplings, the requirement of $N=2$ supersymmetry fixes completely its form
giving the following Lagrangian~\cite{SEIWI}
\beq
L = \frac{1}{4 \pi} Im \left\{ \tau (A) \left[ \partial_{\mu} \bar{A} 
\partial A - \frac{1}{4}\left( F^2 - \frac{i}{2}\epsilon^{\mu \nu \rho 
\sigma} F_{\mu \nu} F_{\rho \sigma} \right) +  Fermions \right] \right\}
\label{genela}
\eeq
where
\beq
\tau (A) = \frac{\partial^2 {\cal{F}} }{\partial A^2}
\label{tau5}
\eeq
is given in terms of an arbitrary function ${\cal{F}} ( A )$ of the
scalar field $A$. This means that the low-energy dynamics
is completely determined by the function ${\cal{F}}$ that in general will
receive both perturbative and non-perturbative contributions.

Comparing eq. (\ref{genela}) with eq. (\ref{newneq2}) we see that at the tree 
level the function ${\cal{F}}$ is given by
\beq
\label{clala}
{\cal{F}}_{cl} = \frac{1}{2} \tau_{cl} A^2 \hspace{1cm} \tau_{cl}=
\frac{\theta}{2 \pi} + i \frac{4 \pi}{e^2}
\eeq

At one loop one gets~\cite{DIVE,SEIBE} instead
\beq
{\cal{F}}_1 = \frac{i}{2 \pi} A^2 \log \frac{A^2}{\mu^2}
\label{1loobeta}
\eeq
that is consistent with the $U(1)_R$ and scale anomaly~\cite{DIVE,SEIBE}. 

It can also be shown that higher loops do not give any contribution to 
${\cal{F}}$.
Only non perturbative effects, as for instance instantons, can give 
an additional 
contribution to ${\cal{F}}$~\cite{SEIBE}. In the last few sections of 
these lectures we will review the
arguments of Seiberg and Witten that led to the exact determination of 
${\cal{F}}$.

We conclude this section by remembering that the $N=2$ super Yang-Mills
theory is an asymptotic free theory with a $\beta$-function $\beta (e) =
- \frac{e^2}{4 \pi^2}$ (for $SU(2)$) getting, in
perturbation theory, only contribution from one-loop diagrams~\cite{GRISARU} 
that also generate the function ${\cal{F}}$ given in eq. (\ref{1loobeta}). 
From the sum of the tree and one-loop contributions ${\cal{F}}_{cl} + 
{\cal{F}}_{1}$ one obtains how the running coupling constant varies with the
scale
\beq
\frac{4 \pi}{e^2 (\mu)} + \frac{1}{\pi} \log \frac{a^2}{\mu^2} \equiv
\frac{4 \pi}{e^2 (a)}
\label{runcou}
\eeq
In terms of the renormalization invariant parameter $\Lambda$ the previous 
equation becomes:
\beq
\label{run}
\frac{e^2 (a)}{ 4 \pi} = \frac{ \pi}{\log \frac{a^2}{ \Lambda^2}}
\hspace{1cm} ; \hspace{1cm} \Lambda^2 = \mu^2 e^{- \frac{4 \pi^2}{g^2 (\mu)}}
\eeq
showing that, because of asymptotic freedom, perturbation theory is good when
$a$ is large.

Many of the previous results as the existence of a manifold of inequivalent 
vacua and the existence of monopole and dyon solutions are also valid for the 
$N=4$ super Yang-Mills theory. This theory, being free from ultraviolet
divergences~\cite{MANDE,BRINK}, has a vanishing $\beta$-function 
and no chiral anomaly~\cite{SOHNIUS}.

\section{Susy algebra in $N=2$ super Yang-Mills}
\label{centra}

The $N=2$ algebra with central charges given in eqs. (\ref{sualgcc}) and
(\ref{sualgcc2}) can be rewritten in four-dimensional notations obtaining:
\beq
\{ Q^{i}_{A} , {\bar{Q}}^{j}_{B} \} = 2 \gamma^{\mu}_{AB} P_{\mu} \delta^{ij}
- 2 \gamma_{5} \epsilon^{ij} V + 2i \epsilon^{ij} \delta_{AB} U
\label{dicech}
\eeq
where
\beq
2 U = - Im Z \hspace{3cm} 2 V = - Re Z
\label{condi}
\eeq

Olive and Witten~\cite{WO} have explicitly computed the central charge in 
the $N=2$ super Yang-Mills theory obtaining:
\beq
U = \int d^3 x \partial_{i} \left[ S^a E_{i}^{a} + P^{a} B^{a}_{i} \right]
\label{cecha}
\eeq
and
\beq
V = \int d^3 x \partial_{i} \left[ P^a E_{i}^{a} + S^{a} B^{a}_{i} \right]
\label{cecha1}
\eeq
where $S = - A_5$ and $P=A_4$.

From the algebra in eq. (\ref{dicech}) applied to a state in the center of 
mass frame where $P_{\mu} = ( M, \vec{0} )$ and from positivity it follows that
\beq
M \geq \sqrt{U^2 + V^2 }
\label{bpscondi}
\eeq

In the asymptotic vacuum given by:
\beq
< P^{a} > =0 \hspace{2cm} < S^{a}> = \sqrt{2} a \delta^{a3}
\label{vacuu}
\eeq
eq. (\ref{bpscondi}) implies
\beq
M \geq \sqrt{2} |a| \sqrt{q^2 + g^2} = \sqrt{2} |a| | q+ ig|
\label{bpscondi2}
\eeq
where $q$ and $g$ are respectively the electric and magnetic charge of the
state. 

Eq. (\ref{bpscondi2}) is the quantum generalization of the BPS condition
found at the classical level. It is a consequence of the supersymmetry algebra
and has now therefore a quantum status. For the BPS 
states, for which the equality sign holds, it is an exact mass formula valid
in the full quantum theory. For this reason the introduction of an extended
supersymmetry allows one to overcome the difficulty mentioned in the second 
point toward the end of section (\ref{MOdua}).

Similar results are also valid in the $N=4$ super Yang-Mills theory where, 
however, instead of only two we have $12$ central charges~\cite{OSBORN}.

\section{$N=2$ versus $N=4$}
\label{Neq24}

In the previous sections we have seen that supersymmetry is an essential
ingredient for having a dual theory in the sense of Montonen-Olive. Restricting
ourselves to the case of pure Yang-Mills theories without super matter we have
analyzed in some detail the theories with $N=2$ and $N=4$. The theory with $N=2$
has some attractive feature as for instance the fact that the supersymmetry
algebra contains only two central charges, the electric and magnetic charges,
while the $N=4$ theory has many more central charges. It has also the nice 
property that the algebra is left unchanged under a simultaneous chiral
transformation of the supercharges and a duality transformation acting on the
electric and magnetic charges as in eq. (\ref{duacomple3}). 

On the other hand it is known from the work of Ref.~\cite{OSBORN} that the
the monopole solution of the $N=2$ theory belongs to the hypermultiplet that
does not contain a spin $1$, while the $W$-bosons, that have a spin $1$, belong
to the $N=2$ chiral multiplet. This shows immediately that the monopoles and
the $W$-bosons cannot be dual in the sense of Montonen-Olive ruling out the
$N=2$ theory.

This problem is overcome in $N=4$ because this theory contains a unique short
multiplet containing one state of spin $1$, four states of spin $1/2$ and five
states with spin $0$ and both the $W$-bosons and the monopoles belong to it.
Therefore this selects the $N=4$ theory as the only theory (without super 
matter)
that can be dual in the sense of Montonen and Olive~\cite{OSBORN}. This theory
has also the attractive feature of being free from ultraviolet divergences.
Therefore its $\beta$-function is vanishing and the gauge coupling constant
is not renormalized. That means that no discussion is needed on which coupling
(the renormalized or the unrenormalized) satisfies the Dirac quantization 
condition~\cite{ROSSI}. Because of this, in the following, we will restrict 
ourselves to this theory
when discussing duality in the sense of Montonen and Olive. We will see, 
however, that duality in the sense of providing different parametrizations
of the same theory will also play an important role in the $N=2$ theory.

\section{SZ quantization condition and the charge lattice}
\label{SZ}

The Dirac quantization condition in eq. (\ref{dirqua}) is only valid for 
particles having either an electric or a magnetic charge. It has been 
generalized by Zwanziger to the case of particles having
both an electric and a magnetic charge. In this case one obtains the so called 
Schwinger-Zwanziger(SZ)~\cite{SCHWINGER,ZWANZI} quantization condition, that 
for two particles with electric and magnetic charges given by $(q_1 , g_1)$ 
and $(q_2 , g_2)$ reads:
\beq
q_1 g_2 - q_2 g_1 = 2 \pi \hbar \, n \hspace{2cm} n= 0, \pm 1 , \pm 2 \dots
\label{dsz}
\eeq

Unlike the Dirac quantization condition, that is only invariant under discrete
duality, the SZ quantization condition is invariant under the full 
electromagnetic duality that acts on $q+ig$ as in eq. (\ref{duacomple3}).
Consequently, without loss of generality, we can assume that there exists a 
subset of purely electric states.

Then we assume also that
\begin{enumerate}

\item{The SZ quantization condition is satisfied.}

\item{The electric and magnetic charges are conserved and any sum of physical
charges is also a physical charge.}

\item{The TCP-theorem is valid and therefore for each particle with 
charges $(q,g)$ there exists also a particle with opposite charges 
$(-q , -g )$.}

\item{There exists at least a particle with non zero magnetic charge.}

\end{enumerate}

Applying the SZ quantization condition in eq. (\ref{dsz}) with a purely electric
state together with a state with arbitrary electric and non vanishing magnetic
charge, that we have assumed to exist, we get that 
the allowed charge values of \underline{purely electric} states are
quantized in terms of a fundamental charge $q_0$.

\beq
 q_i  = n_i q_0   \hspace{3cm} n_i =0, \pm 1 , \pm 2 \dots
\label{elchaqua}
\eeq
where $q_0$ is the electric charge of a state that is physically realized
and is independent from the magnetic charge of the state that we have used in
the SZ quantization condition (see Ref.~\cite{GO} for details). 

Applying then the SZ quantization condition to the case of a purely electric
state with electric charge equal to $q_0$ and of an arbitrary state with 
magnetic charge $g_i$ we get that also the magnetic charge is quantized

\beq
g_i = n_i g_0   
\hspace{2cm} n_i =0, \pm 1 , \pm 2 \dots
\label{machaqua}
\eeq
independently of the electric charge of the state, where
\beq
g_0 = \frac{2 \pi \hbar n_0 }{q_0} 
\label{g0q0}
\eeq
$n_0$ is an integer depending on the theory under consideration.

Let us consider now two dyons with the same magnetic charge $g_0$ and with
electric charges equal to $q_1$ and $q_2$ respectively. The SZ quantization
condition implies:
\beq
q_1 - q_2 = \frac{2 \pi \hbar n}{g_0} = \frac{n}{n_0} q_0 = m q_0
\label{dsz1}
\eeq
where $n$ and $m$ are integers.
The last step in the previous equation follows from eq. (\ref{elchaqua})  since
the state with charge equal to $( q_1 - q_2 ,0)$ is a purely electric state.

Eq. (\ref{dsz1}) implies that the electric charges $q_1$ and $q_2$ of particles
with magnetic charge equal to $g_0$ must be equal to
\beq
q = q_0 \left(n + \frac{\theta}{2 \pi} \right)  
\label{solu3}
\eeq
where $\theta$ is an arbitrary real parameter that is, in some sense, an 
angular variable since $ \theta \rightarrow \theta + 2 \pi$ is equivalent to
shifting $n$ by a unit ($n \rightarrow n+1$).

Applying the SZ quantization condition to two particles with charges equal
to $(q_1 , m g_0 )$ and $(q_2 , m g_0 )$ respectively one gets:
\beq
q_1 - q_2 = \frac{2 \pi \hbar n}{mg_0} = \frac{n q_0}{m n_0} = p q_0
\label{dsz2}
\eeq
where $n,m$ and $p$ are all integers. The last step in the previous equation
follows again, as in eq. (\ref{dsz1}), from eq. (\ref{elchaqua}) since the 
state with charge $ ( q_1 - q_2 , 0)$ is a purely electric state. 
 
The most general solution of eq. (\ref{dsz2}) is of the form:
\beq
q = q_0 \left(n + \frac{\theta}{2 \pi} f_m \right)
\label{solu2}
\eeq
where $f_m$ is an arbitrary number depending only on the magnetic charge of 
the particle. It can be determined by  inserting  eq. (\ref{solu2}) in the SZ 
quantization condition applied to two states with charges equal to 
$(q_1 , m g_0 )$ and $(q_2 , g_0 )$ obtaining
\beq
( q_1 - m q_2 ) g_0 =  2 \pi n_0 \left[ n_1 - m n_2 + \frac{\theta}{2 \pi}
( f_m - m f_1 ) \right] = 2 \pi \hbar n
\label{dsz4}
\eeq
This equation is satisfied if we take:
\beq
f_m = m f_1 = m
\label{solu5}
\eeq
since $f_1 =1$ from eq. (\ref{solu3}). 

We conclude therefore that the electric and magnetic charges of a dyon must 
be given by:
\beq
q+i g = q_0 \left( m \tau + n \right)
\label{latt1}
\eeq
with integers $n$ and $m$, where
\beq
\tau = \frac{\theta}{2\pi} + i \frac{ 2 \pi \hbar n_0}{q_{0}^{2}}
\label{tau}
\eeq

Eq. (\ref{latt1}) implies that the charges of a dyon must lie on a 
two-dimensional lattice with periods $q_0$ and $q_0 \tau$. 
This follows in a straightforward way from the assumptions made at the 
beginning of this section. Notice that $Im \tau > 0$.

The quantity $\tau$ contains a parameter $\theta$ that in the gauge theory 
arises when one includes, together with the usual kinetic term, a term 
proportional to the vacuum angle $\theta$ containing the topological charge
density. The formula in eq. (\ref{latt1}) includes also the Witten 
effect~\cite{WITTEN}
since the electric charge of a dyon gets an additional contribution
if $\theta \neq 0$ due to its magnetic charge. In fact, from eq. 
(\ref{latt1}) one gets the following electric charge of a dyon:
\beq
q = q_0 \left(n + \frac{\theta}{2 \pi} m \right)
\label{Witenef}
\eeq

In conclusion, under the assumptions stated at the beginning of this section, 
we have shown that the electric and magnetic charges of an arbitrary state are
given by the expression in eq. (\ref{latt1}). The question now is how to select
those states that are single particle states. This can be done if we restrict
ourselves to BPS saturated states that, as we have seen in sect. 
(\ref{centra}), have a mass given by an exact quantum formula:
\beq
M = \sqrt{2} |a|| q+ ig |
\label{bpssatu}
\eeq

A single particle BPS-saturated state with mass $M$ must be stable and this 
is the case if it cannot decay into a couple of BPS saturated states with
mass $M_1$ and $M_2$, i.e.
\beq
M < M_1 +M_2
\label{stabi}
\eeq
Using for the mass the expression in eq. (\ref{bpssatu}) together with
the exact expression for the charge given in eq. (\ref{latt1}) one can easily
see, by means of the Schwarz inequality, that eq. (\ref{stabi}) is satisfied 
if and only if the integers $(n,m)$ in eq. (\ref{latt1}) are coprimes. This 
implies that
the stable states with zero magnetic charge $(n ,0)$ are only the three 
states with $n=0, \pm 1$; the states with magnetic charge corresponding to 
$m = \pm 1$ are all stable states; the states with magnetic charge 
corresponding  to $m= \pm 2$ are only stable if their electric charge 
corresponds to odd values of $n$; the states with magnetic charge $m = \pm 3$
are stable if $n$ is different from $0$ and is not a multiple of $3$ and so on. 

\section{Riformulation of Montonen-Olive duality}
\label{RIMO}

We are now in a position to riformulate the Montonen-Olive duality in a way in
which the $W$-bosons, the magnetic monopoles and more in general 
all the dyons of the spectrum are treated in a completely democratic 
way~\cite{OLIVE}. We will see that we will not just have
an electric and magnetic description, but we will have an infinite number
of descriptions depending on which states of the charge lattice we are 
choosing as fundamental particles. Having seen that, if we limit ourselves to
super Yang-Mills theories without super matter, the only theory that could
have a duality in the sense of Montonen-Olive is the $N=4$ super Yang-Mills 
theory, in the following we will concentrate on this theory. 

The usual formulation of the $N=4$ super Yang-Mills is obtained by considering
the states with zero magnetic charge and with electric charge equal to 
$\pm q_0$ corresponding to the $W$-bosons that get a mass given by the formula
in eq. (\ref{bpssatu}) through the Higgs mechanism, together with the massless 
states at the origin of the charge lattice having vanishing electric and 
magnetic charges and corresponding to the photon and Higgs particle. Selecting
these states we have
determined one of the periods of the lattice. The other period is also fixed
when we specify the value of the vacuum angle $\theta$. 
We then ascribe a short $N=4$ supermultiplet to each of the three states with 
charge equal to $0$, $q_0$ and $- q_0$ and, having fixed the value of $\theta$,
we can explicitly write the full Lagrangian of $N=4$ super Yang-Mills 
containing only the states of the lattice that we have chosen. If the theory
is dual in the sense of Montonen-Olive the other stable states of the charge
lattice must appear as solitons or bound states of solitons.

On the other hand if the theory is dual in the sense of Montonen-Olive one
could also start from another couple of stable states of the charge lattice
corresponding to a certain dyon of the theory with a complex charge given by 
$ \pm q_{0} '$ and with mass equal to $M = \sqrt{2} |a| | q_{0} '|$, together 
with the 
massless photon and Higgs states located at the origin of the charge lattice
and specify the vacuum angle $\theta$ by giving another vector 
$ q_{0} ' \tau ' $ of the lattice that is not aligned with $q_0 '$. We can 
again ascribe a $N=4$ short multiplet
to any of the states previously chosen and write, as before, a $N=4$ super
Yang-Mills Lagrangian containing the states with charges equal to $0$ and 
$ \pm q_0 '$ and with a specified vacuum angle $\theta$.  Also in this case
the remaining stable states of the charge lattice will show up as solitons
or bound states of solitons of the new Lagrangian. Duality 
in the sense of Montonen and Olive means that all the theories
based on any pair of independent vectors of the  charge lattice are 
equivalent.

Since the vectors $q_0 '$ and $ q_0 ' \tau '$ form an alternative basis of the
charge lattice it must be possible to express them  in terms of the original 
vectors $q_0$ and $ q_0 \tau$ through the relation:
\beq
q_0 ' \tau ' = a q_0 \tau + b q_0 \hspace{2cm} 
q_0 '  = c q_0 \tau + d q_0 
\label{modtra}
\eeq
with $a,b,c$ and $d$ integer numbers. 

Since it must also be possible to express $q_0 $ and $q_0 \tau$ in terms
of $q_0 '  $ and $q_0 '  \tau '$ the integers of the transformation must 
satisfy the equation:
\beq
ad - bc =1
\label{deteq1}
\eeq

Therefore the transformations from a basis to another basis form the modular
group $SL(2, Z)$.

Eqs. (\ref{modtra}) imply a relation between $\tau$ and $\tau '$ given by
\beq
\tau ' = \frac{a \tau + b}{c \tau + d}
\label{modtra1}
\eeq
that provides a connection between the values of the parameters 
$( \theta, q_0)$ in the two choices of basis vectors and actions.

The modular group is generated by the two transformations:
\beq
T \hspace{1cm}: \hspace{1cm} \tau \rightarrow \tau +1 \hspace{1cm}
\rightarrow \hspace{1cm} \theta \rightarrow \theta + 2 \pi
\label{ti}
\eeq
that is a symmetry of the theory because the physics is periodic when we
translate $\theta$ by $2 \pi$, and
\beq
S \hspace{1cm}: \hspace{1cm} \tau \rightarrow - \frac{1}{\tau}  \hspace{1cm}
\rightarrow \hspace{1cm} q_0  \rightarrow \frac{2 \pi \hbar n_0}{q_0} 
\hspace{1cm}(\,\, if \,\, \theta=0)
\label{si}
\eeq
that relates weak coupling with strong coupling (compare
with eq. (\ref{duality3}), $n_0 =2$). 

The mass of the BPS-saturated states of the theory is proportional to the 
absolute value of the charge
\beq
M \sim | q+ ig| = | q_0 ( m \tau + n)|
\label{massbps}
\eeq
and is left invariant if we transform $\tau$ as in eq. (\ref{modtra1})
and $q_0$ and the charge vector $\left( \begin{array}{c} m \\
                        n  \end{array} \right)$ as follows
\beq
q_0 \rightarrow q_0 ' = q_0 ( c \tau + d) \hspace{1cm}
\left( \begin{array}{c} m \\
                        n  \end{array} \right) \rightarrow
                       \left( \begin{array}{c} m' \\
                                               n'  \end{array} \right) =
\left( \begin{array}{cc} d & -c \\
                 -b &  a   \end{array} \right)
  \left( \begin{array}{c} m \\
                        n  \end{array} \right)
\label{modtrasf}
\eeq
with $ad-bc=1$.

The modular group does not only perform a transformation from a system of 
basis vectors to another one, but acts also on the integer charge vector
$\left( \begin{array}{c} m \\
                        n  \end{array} \right)$ rotating it into a new integer
charge vector $\left( \begin{array}{c} m' \\
                        n'  \end{array} \right)$.
In other words a modular transformation transforms $q+i g$ expressed in terms 
of the basis vectors $q_0$ and $q_0 \tau$ and of the integers $n$ and $m$ into
an expression having the same form in terms of the new basis vectors $q_0 '$ 
and $q_0 ' \tau '$ and of the new integers $n'$ and $m'$ related to the old 
ones by eqs. (\ref{modtra1}) and (\ref{modtrasf}). The invariance under the 
modular group requires that the existence in the spectrum of a state with a 
certain pair of integers implies also the existence in the spectrum of the
state with other integers obtained from the first ones by the action of a
modular transformation as in the second equation of  (\ref{modtrasf}).
 
In particular from  eq. (\ref{modtrasf}) it follows that, given the
existence in the spectrum of the $W^{+}$-boson corresponding to $m=0$ and $n=1$,
the invariance under the modular group implies also the existence of the
transformed state:
\beq
\left( \begin{array}{c} 0 \\
                        1  \end{array} \right) \rightarrow
                       \left( \begin{array}{c} -c \\
                                               a  \end{array} \right) =
\left( \begin{array}{cc} d & -c \\
                 -b &  a   \end{array} \right)
  \left( \begin{array}{c} 0 \\
                        1  \end{array} \right)
\label{modtrasf1}
\eeq
Since the condition $ad-bc=1$ is equivalent to require that $c$ and $a$ are 
coprimes, the existence of the $W^+$-boson implies the existence in the
spectrum of all stable states of the charge lattice as discussed at the end of
the previous section. This is a direct consequence of the Montonen-Olive 
duality.

Let us consider the states with $c=-1$. They are of the type $ \left( 
                             \begin{array}{c}  1 \\
                                               a  \end{array} \right) $
where $a$ is an arbitrary integer. These are the dyons in eq. (\ref{elema}).
The next case is $c =-2$. In this case we expect the existence of the states
$ \left( \begin{array}{c}                      2 \\
                                               a  \end{array} \right) $
where $a$ is odd. The existence of such states was shown by Sen~\cite{SEN}.
Evidence for the existence of stable states with higher values of $c$ can be 
found in Ref.~\cite{PORRATI}.
 
\section{Global parametrization of moduli space in $N=2$ theory}
\label{global}

In the last few sections we will be shortly describing the beautiful 
paper of Seiberg and Witten~\cite{SEIWI} where an exact expression for 
$\tau (A)$ (see eq. (\ref{genela})) in the low energy effective action of the 
$N=2$ super Yang-Mills theory has been constructed.

Unlike the $N=4$ theory which can be equivalently formulated either in 
terms of the original fundamental fields or in terms of the monopoles or
more in general of the dyons of the single particle spectrum with essentially 
the same Lagrangian, the $N=2$ theory cannot satisfy the Montonen-Olive duality
because the fundamental fields and the magnetic monopoles belong to two 
different $N=2$ superfields. The first are in the chiral $N=2$ vector multiplet
while the magnetic monopoles and dyons are in the hypermultiplet~\cite{OSBORN}.

Nevertheless the $N=2$ theory can be formulated either in terms of the
variables $A, A_{\mu}$ and $\tau(A)$, as we have done in eq. (\ref{genela}),
or in terms of the dual variables $A_D , A_{D \,\mu}$ and $\tau_{D} (A_D )$ 
in pretty much the same way that free electromagnetism can be formulated
either in terms of the vector potential $A_{\mu}$ related to the field strenght
by $F_{\mu \nu} \equiv \partial_{\mu} A_{\nu} - \partial_{\nu} A_{\mu}$  or in
terms of the dual vector potential $A_{D \, \mu}$ related to the dual field 
strenght by ${}^* F_{\mu \nu} = \partial_{\mu} A_{D\, \nu} - \partial 
A_{D\, \mu}$.

In order to explain this let us first summarize some general property of the
$N=2$ super Yang-Mills theory. 
 
In sect. (\ref{semi}) we have seen that in the $N=2$ theory the
low energy effective theory is completely fixed by giving a holomorphic
function ${\cal{F}}(A)$. In terms of ${\cal{F}}$ we can construct 
the K{\"{a}}hler potential:
\beq
K ( A , {\bar{A}} ) = Im \left( \frac{\partial {\cal{F}}}{\partial A } 
{\bar{A}} \right)
\label{Kae}
\eeq  
and the metric
\beq
(ds)^2 = \frac{\partial}{\partial A} \frac{\partial}{ \partial {\bar{A}}}
K(A, {\bar{A}}) dA d{\bar{A}}= Im ( \tau (A) ) dA d{\bar{A}}
\hspace{1cm} \tau (A) = \frac{\partial^2 {\cal{F}} (A)}{\partial A^2} 
\label{met}
\eeq
 
We have seen that the moduli space of the $N=2$ theory is in the semiclassical
theory parametrized by the vacuum expectation value of the scalar field that
we have denoted by the complex number $a$. However $a$ cannot provide a
global description of the moduli space. In fact the metric $Im (\tau (a) )$, 
that is a positive definite harmonic function divergent for $|a| \rightarrow
\infty$, must have a minimum. But a globally defined harmonic function
cannot have a minimum and consequently the
variable $a$ cannot provide a global parametrization of the moduli space.

Therefore in Ref.~\cite{SEIWI} it was proposed to choose the gauge invariant
quantity $u = \frac{1}{2} Tr ( \Phi^2 )$ as the one that provides a global
parametrization of the moduli space and to regard both $a (u)$  and the dual 
variable $a_D (u) \equiv \frac{\partial {\cal{F}}}{\partial a}$ as functions 
of $u$. In terms of both $a$ and $a_D$ the metric in eq. (\ref{met}) assumes
the form
\beq
(ds)^2 = Im \left( \frac{d a_D}{ d a} da d{\bar{a}} \right)=
Im \left( da_D d {\bar{a}} \right) = - \frac{i}{2} \left[ da_D d{\bar{a}}
- da d{\bar{a}}_D \right]
\label{newme}
\eeq
that is symmetric under the exchange $a \leftrightarrow a_D$.

Introducing the vector $v^{\alpha} = \left( \begin{array}{cc} a_D \\ a  
\end{array}\right)$ we can rewrite the metric in the more compact form:
\beq  
(ds)^2 = - \frac{i}{2} \epsilon_{\alpha \beta} \frac{d v^{\alpha}}{du}
\frac{d {\bar{v}}^{\beta}}{d{\bar{u}}} du d{\bar{u}}
\label{nmet}
\eeq 
that clearly show its invariance under the transformation:
\beq
v \rightarrow M v + c
\label{sl2r}
\eeq
where $M$ is a matrix of $SL(2,R)$ and $c$ is a constant vector.

An arbitrary matrix of $SL(2,R)$ is generated by the action of two independent
matrices $T_b$ and $S$. The first one $T_b$
\beq
T_b = \left(  \begin{array}{cc} 1 & b \\
                                0 & 1  \end{array} \right)
\label{tb}
\eeq
leaves $a$ invariant and transforms $a_D$ according to
\beq
a_D \rightarrow a_D + b a
\label{tbtra}
\eeq
This implies that $\tau (a)$ is just translated
\beq
\tau (a) \rightarrow \tau (a) + b
\label{tautra}
\eeq
resulting in a translation for the vacuum angle $\theta$
\beq
\theta \rightarrow \theta + 2 \pi b
\label{thetra}
\eeq
Since physical quantities are invariant when
\beq
\theta \rightarrow \theta + 2 \pi n
\label{thetra1}
\eeq
for any integer $n$, comparing eqs. (\ref{thetra}) and (\ref{thetra1}) we
deduce that $b=1$ and consequently that the transformation associated to the
matrix $T_{b=1}$ is a symmetry of the theory. By selecting $b=1$ we have
reduced the original $SL(2,R)$ symmetry group to $SL(2,Z)$.

The other independent generator 
\beq
S = \left( \begin{array}{cc} 0 & 1 \\
                             -1 & 0 \end{array} \right)
\label{stra}
\eeq
does not correspond to a symmetry of the theory, but provides a transformation
between two different parametrizations of the theory. In fact the low energy 
effective Lagrangian can be represented either in terms of the variables
$( A^{\mu}, \lambda, A; \tau (A)) $ or in terms of the dual ones 
$( A^{\mu}_D, \lambda_D, A_D; \tau_D (A_D ) =
-1/ \tau (A) ) $. In order to more clearly see the relation between the two 
formulations it is convenient to set the vacuum angle $\theta=0$. Then we see 
that, if
$Im \tau (A)= \frac{4 \pi}{g^2}$, then $Im \tau_D (A_D )= \frac{g^2}{4 \pi}$.
Therefore one description may be more suitable for weak coupling, while the
other for strong coupling.

In the final part of this section we discuss the exact mass formula proposed
in Ref.~\cite{SEIWI} for the BPS saturated states in the $N=2$ theory. At the 
semiclassical level
the mass of the BPS saturated states is given in eq. (\ref{mass4}). 
Noticing that the coefficient of $n_m$ in eq. (\ref{mass4}), with a suitable
choice of $C$, is equal to $a_D$ eq. (\ref{mass4}) can be rewritten as follows
\beq
Z = a n_e + a_D n_m = \left( \begin{array}{cc} n_m & n_e \end{array} \right)
\left( \begin{array}{cc} a_D \\
                          a \end{array} \right) \hspace{2cm} M= \sqrt{2} |Z|
\label{mass4e}
\eeq

Seiberg and Witten~\cite{SEIWI} proposed eq. (\ref{mass4e}) as an exact
formula for the BPS states and made several checks for confirming its 
validity.

In particular $Z$ is invariant under the transformation
\beq
\left( \begin{array}{cc} a_D \\
                          a \end{array} \right)
\rightarrow M \left( \begin{array}{cc} a_D \\
                          a \end{array} \right)
\hspace{2cm}
\left( \begin{array}{cc} n_m & n_e \end{array} \right) \rightarrow
\left( \begin{array}{cc} n_m & n_e \end{array} \right) M^{-1}
\label{modinv7}
\eeq
where $M$ is a matrix of $SL(2,Z)$ because the vector
$\left( \begin{array}{cc} n_m & n_e \end{array} \right)$ has integer
entries and its transformed must also have integer entries. This is an 
independent way to derive the reduction of $SL(2,R)$ to $SL(2,Z)$.
Actually this procedure forces also the extra parameter $c$ in eq. 
(\ref{sl2r}) to be equal to zero.

\section{Singularity structure of moduli space}
\label{singu}

In this section we study the singularity structure of $a$ and $a_D$ as 
functions of the variable $u$, that provides a global parametrization of the 
moduli space. 

In the semiclassical region corresponding to a large value of $u$ we get 
\beq
\label{5.1}
a = \sqrt{2u} \hspace{2cm} a_D = i \frac{\sqrt{2u}}{\pi} \left[ 2 \log 
\frac{\sqrt{u}}{\Lambda} + 1 \right]
\eeq
Under a rotation around $u = \infty$ given by
\beq
\label{5.2}
\log u \rightarrow \log u + 2i \pi
\eeq
$a$ and $a_D$ are not monodromic functions, but transform according to
\beq
\label{5.3}
a \rightarrow - a \hspace{2cm} a_D \rightarrow - a_D + 2 a
\eeq

It is interesting that the asymptotic freedom property of the theory
is responsible for this monodromy transformations.

The existence of a singularity requires the existence of at least another
singularity. But, if we had only one additional singularity, it is easy to 
see that $a$ would have been a good global parameter being the monodromy 
group an abelian group. Since this is not possible we must require 
the existence of at least two additional singularities. 

Following the example of what is happening in some $N=1$ supersymmetric
theories Seiberg and Witten assume that the singularities occur at those
points of the moduli space where additional massless particles appear in the
spectrum. In the classical theory this occurs for $a=0$ where the $SU(2)$
symmetry is restored and $W^{\pm}$ become massless. They bring good 
indications against this possibility in the quantum theory  and instead
choose the singularities at the points $u_0$  where the monopole 
with  $(n_m ,n_e )= (1,0 )$ and $u_0 '$  where the dyon with $(n_m , n_e ) 
= (1, -1)$ become massless.

Using the exact formula in eq. (\ref{mass4e}) it is easy 
to see that this occurs
when $a_D ( u_0 ) =0 $ with $a (u_0 ) \neq 0$  and when $a_D ( u_0 ') - 
a (u_0 ' ) =0 $ with $a (u_0 ' ) , a_D ( u_0 ') \neq 0$ respectively.
The existence of a $Z(2)$ symmetry that transforms $u$ in $- u$ suggests to
choose $u_0 ' = - u_0$. They introduce a new dimensionless variable $u$ obtained
from the previous one by dividing it with the square of a mass parameter 
chosen in such a way that the singularity due to the vanishing of the mass
of a magnetic monopole occurs at $u =1$. Then by the $Z (2)$  symmetry  the 
other singularity due to the dyon occurs at $u =-1$. 

The monodromy around the singularity at $u=1$ can be easily computed by 
observing that the low energy theory at the point $u=1$ consists of a
"magnetic" $N=2$ super QED (the matter has magnetic and non electric 
charge). This theory is not asymptotically free and the coefficient of the
$\beta$-function, besides a sign, has a factor $1/2$ of difference with  
respect to the $\beta$-function previously used for studying the singularity
around $u = \infty$. By taking into account this difference in the 
$\beta$-function one arrives at the following monodromy transformations
around the point $u=1$:
\beq
\label{5.4}
a_D \rightarrow a_D \hspace{2cm} a \rightarrow a - 2 a_D
\eeq
Finally the monodromy around the point $u= -1$ must be consistent with the
previous ones and one obtains
\beq
\label{5.5}
a_D \rightarrow - a_D + 2a \hspace{2cm} a \rightarrow - 2 a_D + 3 a
\eeq

\section{Explicit solution}
\label{exsolu}
Having established the singularities and the monodromy transformations of 
$a$ and $a_D$ Seiberg and Witten were able to construct an explicit solution 
that is given by
\beq
a(u) = \frac{\sqrt{2}}{\pi} \int_{-1}^{1} dx \frac{\sqrt{x-u}}{\sqrt{x^2 -1}}
\hspace{2cm}
a_D (u) = \frac{\sqrt{2}}{\pi} \int_{1}^{u} dx 
\frac{\sqrt{x-u}}{\sqrt{x^2 -1}}
\label{6.1}
\eeq
and that is singular at the points $u= \pm 1, \infty$ with the right 
monodromies.

In terms of the previous functions one can construct the coefficient of the
kinetic term of the gauge field
\beq
\tau (u) = \frac{\frac{d a_D}{du}}{\frac{da}{du}}
\label{6.2}
\eeq
that satisfies the important property : $Im \tau > 0$ for any $u$.

The classical vacuum degeneracy is not lifted by quantum effects not even
after having taken into account  the non-perturbative effects!!

\vskip 1.0cm

{\large {\bf {Acknowledgements}}}
\vskip 0.5cm
I thank Lorenzo Magnea for a critical reading of the manuscript.
This research was partially supported by the EU, within the framework
of the program ``Gauge Theories, Applied Supersymmetry and Quantum Gravity''
under contract SCI-CT92-0789.

\end{document}